%% file: pap3.tex
\documentclass[11pt]{article}
\usepackage{epsfig,sint,macros,cite}
\usepackage{amssymb}
\begin{document}
\input title.tex
\input sect1.tex

\input sect2.tex

\input sect3.tex

\input sect4.tex

\input sect5.tex 
\input sect6.tex

\input sect7.tex

\input ack.tex
\begin{appendix}
\input appendix1.tex
\end{appendix}
\bibliography{biblist}        
\bibliographystyle{h-elsevier}   
\end{document}

%% file: title.tex
\begin{titlepage}
\begin{flushright}
  DESY-99-075\\
  Edinburgh 6/99\\
  OUTP-99-27P\\
  UTCCP-P-67
\end{flushright}

\vskip 0.5 cm
\begin{center}
  {\Large\bf 
  Precision computation of the strange quark's mass  \\[0.5ex]
  in quenched QCD\\[0.5ex] }
\end{center}
\vskip 0.5 cm
\begin{center}
{\large Joyce Garden$^{\scriptscriptstyle a}$, 
     Jochen Heitger$^{\scriptscriptstyle b}$,
     Rainer Sommer$^{\scriptscriptstyle b}$ and
     Hartmut Wittig$^{\scriptscriptstyle c,d,}$\footnote{PPARC Advanced Fellow}
\vskip 0.5cm
(ALPHA and UKQCD Collaborations)
\vskip 0.5 cm
$^{\scriptstyle a}$
Department of Physics \& Astronomy, University of Edinburgh\\
Edinburgh EH9~3JZ, Scotland
\vskip 2.5ex
$^{\scriptstyle b}$
DESY-Zeuthen \\
Platanenallee 6, D-15738 Zeuthen, Germany
\vskip 2.5ex
$^{\scriptstyle c}$
Theoretical Physics, University of Oxford \\
1~Keble Road, Oxford OX1~3NP, UK
\vskip 2.5ex
$^{\scriptstyle d}$
Center for Computational Physics, University of Tsukuba \\
Tsukuba, Ibaraki 305-8571, Japan
\vskip 1.0cm
{\bf Abstract}}
\vskip 0.7ex
\end{center}

We determine the renormalization group invariant quark mass
corresponding to the sum of the strange and the average light quark
mass in the quenched approximation of QCD, using as essential input
the mass of the K-mesons. In the continuum limit we find
$(M_{\strange} +\Mlight)/\Fk=0.874(29)$, which includes systematic
errors. Translating this non-perturbative result into the running
quark masses in the $\msbar$-scheme at $\mu=2\,\GeV$ and using the
quark mass ratios from chiral perturbation theory, we obtain
$\mbar_{\strange}(2\,\GeV)=97(4)\,\MeV$. With the help of recent
results by the CP-PACS Collaboration, we estimate that a 10\% higher
value would be obtained if one replaced $\Fk$ by the nucleon mass to
set the scale. This is a typical ambiguity in the quenched
approximation.

\vfill

\begin{center}
June 1999
\end{center}

\eject

\vfill

\eject

\end{titlepage}

%% file: sect1.tex
\section{Introduction \label{s_intro}}

Quark masses are fundamental parameters of the standard model, which
have to be determined from experimental observations confronted with
theoretical predictions~\cite{reviews:quarkmasses}. At present the
most precise theoretical predictions which allow for the
determination of ratios of the three light quark masses are based on
chiral perturbation theory~\cite{leutwyler:94}. A detailed analysis
yielded~\cite{leutwyler:1996}
\bes
 M_\up/M_\down = 0.55\pm0.04\,,\quad 
 M_\strange/\Mlight = 24.4 \pm1.5  
 \label{e_ratios}
\ees
with
\bes
 \Mlight = \frac12(M_\up+M_\down) \,.
\ees
Unlike these ratios, the overall magnitude of the quark masses is
not accessible to chiral perturbation theory combined with
experimental data alone. One should therefore determine a particular
linear combination of quark masses by comparing lattice QCD
predictions~\cite{quark:APE1,quark:APE2,quark:APE3,hadr:ukqcd,quark:gupta,quark:gough,impr:qcdsf,impr:roma2,quark:marti,quark:SESAM,quark:dwf,quark:jlqcd_ks,qspect:CPPACS}
or QCD sum
rules~\cite{quark:bijnens,quark:jamin1,quark:jamin2,quark:narison,quark:chetyrkin,quark:colangelo
,quark:prades,quark:yndurain,quark:dosch,quark:lellouch,quark:maltman} to
experiments.

In this work we use the masses of the K-mesons and a computation in
the quenched approximation to QCD to determine
$M_{\strange}+\Mlight$. Our analysis employs the $\Oa$ improved
lattice theory, the quark mass is renormalized completely
non-perturbatively~\cite{mbar:pap1} and the continuum limit is taken
(with a rate proportional to $a^2$). Hence, with respect to the last
two points, it improves on many previous calculations.

In general quark masses are scale- and scheme dependent quantities.
It is therefore desirable to compute the {\em renormalization group
invariant quark masses}, which -- being both scale- and
scheme-independent -- are naturally taken as fundamental parameters
of QCD. We recall that they are defined in terms of the high energy
behaviour of the running masses $\mbar(\mu)$:
\bes
 M_i &\equiv& \lim_{\mu\to\infty} \left\{(2b_0\gbar^2(\mu))^{-d_0/2b_0}\, 
 \mbar_i(\mu)\right\} \,, \label{e_M_i1}\\
 &&b_0=11/(4\pi)^2,\; d_0=8/(4\pi)^2 \,.
\ees
On the other hand, the renormalization group invariant masses
$M_i$ are related to 
the bare current quark masses $m_i$ by
\bes
 M_i  = \zM m_i 
  \label{e_M_i2}\,
\ees
with a (flavour independent) renormalization factor $\zM$ which was 
recently  computed by the ALPHA Collaboration \cite{mbar:pert,mbar:pap1}.
This non-perturbative result is the basis of our present 
calculation.\footnote{Since the bare mass is involved, $\zM$
depends on the regularization. The complete calculation of 
\cite{mbar:pap1} was done in $\Oa$ improved lattice QCD, which we use here as
well.}
The renormalization problem and its solution were
discussed in 
detail in references \cite{mbar:pert,mbar:pap1}. 
The current quark masses themselves, are defined through the
PCAC relation,
\bes
 \partial_\mu A_\mu(x) =
  (m_i + m_{j}) P(x),
  \label{e_PCAC}
\ees
in terms of the axial current,
\bes
 A_\mu(x) = \psibar_{i}(x) \gamma_\mu \gamma_5 \psi_{j}(x),
\ees
and the {\ps} density,
\bes
 P(x) = \psibar_{i}(x)  \gamma_5 \psi_{j}(x) \, .
\ees
Applied to the vacuum-to-K matrix elements
it reads
\bes
 M_{\strange} + \Mlight = \zM (m_{\strange}+ \mlight) 
 = \zM {F_{\rm K} \over G_{\rm K}} m^2_{\rm K}\,,
 \label{e_fund}
\ees
where $\Fk$ is the K-meson decay constant, and $G_{\rm K}$ denotes
the vacuum-to-K matrix element of the {\ps} density.\footnote{Our
convention is that $\Fk,\Gk$ denote the matrix elements of the bare
operators. Renormalization factors are written explicitly. For the
$\Oa$ improved lattice theory they can be found in the appendix.}
\Eq{e_fund} is the fundamental relation which we shall exploit in
this work. It is used in the following way: we compute 
$\zM {F_{\rm K} \over G_{\rm K}}$ and multiply with the experimental
squared mass,
\bes
m^2_{\rm K} = \frac12(m^2_{\rm K^+} + m^2_{\rm K^0})_{\rm QCD} = 
 (495\,\MeV)^2 \,, \label{e_m_K}
\ees
to obtain $M_{\strange} + \Mlight$. By the subscript ``QCD'' we
indicate that we have used the masses in pure QCD with
electromagnetic interactions switched off, since obviously the
lattice QCD result is valid for a world where $\alpha_{\rm
em}=0$. In practice this is achieved by subtracting an estimate of
the electromagnetic effects from the experimental numbers. The
numerical estimate in \eq{e_m_K} was obtained from Dashen's theorem
\cite{Dashen} being well aware that the accuracy of this estimate
may be only around 0.5\% in $m^2_{\rm K}$ \cite{em_corr}.

The combination $\zM {F_{\rm K} \over G_{\rm K}}$ carries the
dimension of an inverse mass. Therefore it is necessary to choose
another dimensionful observable to form a dimensionless ratio which
has a continuum limit (this is often called ``setting the
scale''). Choosing $\rnod$ \cite{pot:r0} or $\Fk$ for this second
observable and extrapolating to the continuum limit yields the
results quoted in the abstract.

In the following we shall first explain our strategy to deal with
some technical difficulties in the computation of ${F_{\rm K} /
G_{\rm K}}$. Since it involves a computation of the ratio $\Fp/\Gp$
for mass-degenerate mesons, we discuss in~\sect{s_ndg} the
dependence of various observables on the difference of quark masses
$m_i-m_j$. In \sect{s_fse} we show that finite size effects are
negligibly small in our calculation. The main result for the quark
mass is presented in \sect{s_cont}, where, among other issues, the
extrapolations to the continuum are discussed. We then proceed to
estimate the {\em ambiguity} which originates from the fact that the
{\em quenched approximation} can not be expected to describe the
real world properly. As a byproduct, we present the calculation of
the kaon decay constant and masses in the vector channel in the
continuum limit. We finish with a discussion of our results and some
comments on how the method may be extended to determine coefficients
of the chiral Lagrangian.

%% file: sect2.tex
\hyphenation{mass-re-nor-ma-li-za-tion}

\section{Strategy \label{s_s}}
\subsection{Chiral perturbation theory}

Let us first recall what chiral perturbation theory can predict
concerning the quark
masses~\cite{chir:GaLe1,chir:GaLe2,leutwyler:94,leutwyler:1996}. Chiral
perturbation theory is based on nothing but the very general
assumption that chiral symmetry is broken spontaneously in the limit
$\Mlight=\Mstrange \to 0$. This allows for a quantitative
description of the pseudoscalar sector in terms of a low-energy
effective Lagrangian, where quark masses appear as ``kinematical
variables'' just like the energy in scattering processes. The
parameters in the chiral Lagrangian are independent of the quark
masses. At order $p^4$, most of the parameters have been determined
by comparison to experimental data (see e.g. \cite{chir:param} for
numerical values). However, the coefficient of the chiral symmetry
breaking quark mass term in the Lagrangian (denoted by $B$ in
\cite{chir:GaLe2}), cannot be determined from experimental data
alone. It can only be fixed when one particular quark mass is known. 
Since all parameters are independent of the quark masses, this may
be done at a convenient reference point. It is important to realize
that this reference point does not have to correspond to a physical
quark mass. The procedure can then be extended to determine other
parameters in the chiral Lagrangian more precisely, once additional
observables are known for suitable quark masses and with
sufficient accuracy. All of this can potentially be achieved by
lattice QCD calculations and can improve predictions such as
\eq{e_ratios}. Also the justification for the truncation of chiral
perturbation theory can and should, of course, be checked.

A particularly convenient way of applying this idea in practice
is as follows. We introduce the ratio
\bes
  \RWI(m_i,m_j) &=& {\Fp \over \Gp} \,,
       \label{e_def_R}
\ees
such that
\bes
 m_i + m_{j} &=& \RWI(m_i,m_j) \, \mp^2(m_i,m_j) \,.
       \label{e_PCAC_l}
\ees 
Defining a reference quark mass
\bes
  \mp(\mref,\mref) = \mk \,,
\ees
with the kaon mass already discussed in the introduction, the 
ratio $\RWI$ for the physical quark masses may be written as
\bes
 \RWI(m_i,m_j) = T(x_i,x_j) \RWI(\mref,\mref)\,, \quad x_i=m_i/\mref
\ees
with a function $T(x_i,x_j)$ which can be computed in chiral
perturbation theory. In particular, using
\cite{chir:GaLe1,chir:param} one finds
\bes 
  T(\mstrange/\mref,\mlight/\mref) \approx 1
  \quad \mbox{and thus}   \quad
  2\mref \approx \mstrange+\mlight \,.
  \label{e_equiv}
\ees
The corrections to the above equation are small, but the overall
uncertainties associated with this statement require a detailed
investigation of the phenomenology. The reason is that the errors of
the parameters in the chiral Lagrangian are correlated, which makes
it non-trivial to estimate uncertainties. Nevertheless we expect
\eq{e_equiv} to be correct to within about 10\%.

In \sect{s_ndg} we shall follow a complementary approach, by
investigating {\ps} meson observables as a function of the average
quark mass and the difference of the quark masses. We find that the
dependence on the latter variable is rather small. It then follows
trivially from our definition of $T$ that \eq{e_equiv} is valid
rather precisely.

Before turning to that investigation let us briefly explain why our
strategy is advantageous. The reasons for quenched QCD and full QCD
are somewhat different but related, having both to do with the
difficulties of extrapolations to 
the region of small quark masses which is at
present inaccessible to direct Monte Carlo calculations.

\subsection{Quenched approximation and full QCD}
The lattice calculations presented in this paper have been performed
in the $\Oa$ improved theory \cite{impr:pap1} with non-perturbative
improvement coefficients \cite{impr:pap3} and mass-renor\-ma\-li\-zation
factor, $\zM$ \cite{mbar:pap1}. Details pertaining to, for instance,
$\Oa$ correction terms in the currents can be found in the appendix.

To motivate our strategy let us first review the straightforward
approach for computing light quark masses. First one assumes isospin
symmetry of the ratios $\RWI$, i.e. effects of $\rmO(\mup-\mdown)$
are neglected, which is well justified given the smallness of this
mass difference. Then one chooses an overall scale, say the
hadronic radius $\rnod$, and determines the bare light quark mass
and strange quark mass such that $r_0 m_{\pi}$ and $r_0 m_{\rm K}$
agree with the experimental numbers. Except for the bare coupling,
all parameters in the QCD Lagrangian are then fixed and the current
quark masses are given by eqs.~(\ref{e_def_R},\ref{e_PCAC_l}), with
\bes
 \langle 0 |A_{\mu}(0)|{\rm K}(p)\rangle =  i p_\mu \Fk\,,
 \quad
 \langle 0 |P(0)|{\rm K}(p)\rangle = \Gk  \,. 
\ees
Here $|{\rm K}(p)\rangle$ denotes a {\ps} state with momentum $p$,
standard infinite volume normalization and the proper flavour
quantum numbers. After renormalization the quark masses can be
extrapolated to the continuum.

While this approach will  ultimately be a clean and straightforward
way to determine the quark masses, it poses two problems in the
quenched approximation.
\begin{itemize} 
 \item The quenched approximation is expected to be misleading for
   very light quark masses such as $\mlight$. A particular
   indication of the failure of the quenched approximation in this
   regime is the presence of logarithmic terms in chiral
   perturbation theory which have no counterpart in the full theory
   \cite{chir:quenched1,chir:quenched2,chir:quenched3}.

 \item With our fermion action and lattice spacings, we cannot
   perform calculations for quark masses which are below
   approximately half of the strange quark mass. For smaller masses,
   the Dirac operator has unphysical zero modes on a certain
   fraction of configurations contributing to the path integral
   (``exceptional configuration'') \cite{impr:pap3,except:fermilab}.
\end{itemize}
The second problem is of a more technical nature. It can in
principle be circumvented by choosing a suitable action, but then
also $\zM$ has to be recomputed. However, the first point must be
taken seriously, since we want to obtain results which are not
misleading with respect to the real theory of interest, full QCD.

It is furthermore advantageous to use the same strategy in full QCD. 
The reason is that it is very likely that for some time to come,
lattice simulations will only reach down to quark masses somewhere
around $m=\mref (\approx \mstrange/2)$. The most precise way to
extrapolate further is then given by chiral perturbation theory as
discussed above.

%% file: sect3.tex
\section{Non--degenerate quarks \label{s_ndg}}

\begin{figure}[tb]
\hspace{0cm}
\vspace{-4.2cm}

\centerline{
\psfig{file=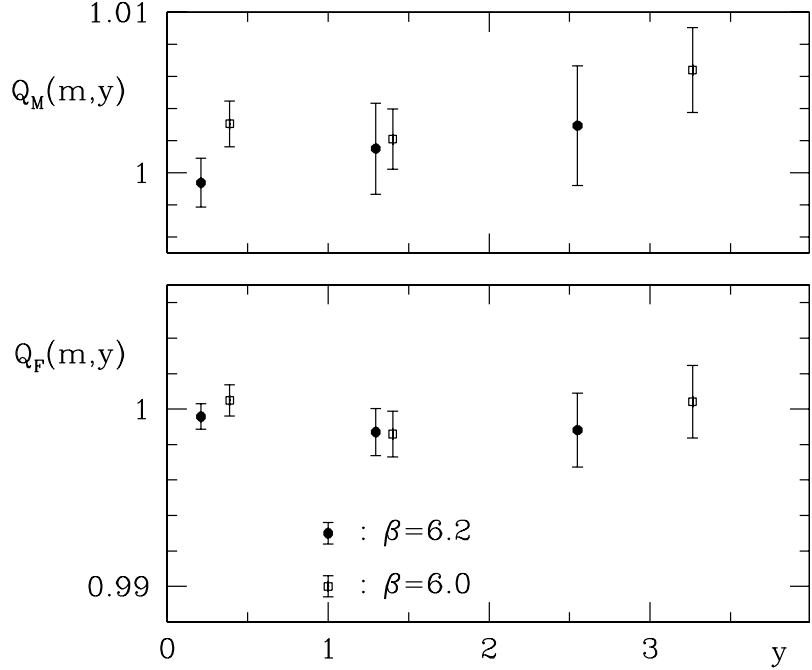,width=14cm}
}
\vspace{-1.2cm}
\caption{\footnotesize
The functions $Q(m,y)$. $Q_{\rm M}$ denotes the case of pseudoscalar masses
and $Q_{\rm F}$ their decay constants. The average of $m_i$ and $m_j$
is between $\approx 1.4\, \mref $ and $\approx 2.4\, \mref$.
\label{f_Q}}
\end{figure}
%
In this section we investigate quantitatively how the {\ps} mass and
decay constant depend on the difference of the two quark masses in
quenched QCD. Apart from some consistency checks which were
performed with the data of the ALPHA Collaboration, the numerical
results presented here were obtained during the course of the UKQCD
simulations \cite{impr:ukqcd_qhad} using the $\Oa$ improved action.

We consider the observables of interest as a function of the average
current quark mass $m=\frac{m_i+m_{j}}{ 2}$ and the parameter $y$
defined by
\bes
   y=(x_i-x_{j})^2 =(m_i-m_j)^2/\mref^2\,, \label{e_y_def}
\ees
with $\mref$ as determined later in \sect{s_cont}. In particular,
we want to show that the functions
\bes
 H_{\rm M}(m,y)=\mp^2\;\; \mbox{and} \;\; H_{\rm F}(m,y)=
  \Fp\,[1+ \ba a\mq]\;
\ees
have little dependence on $y$. In other words, the ratios
\bes
 Q_{\rm M}(m,y)&=&H_{\rm M}(m,y)/H_{\rm M}(m,0)\,,\\
 Q_{\rm F}(m,y)&=&H_{\rm F}(m,y)/H_{\rm F}(m,0)\,,
\ees
are close to unity. Note that the two functions $Q_{\rm M}$ and $T$
are equivalent at the special point $m=\mref$, since $Q_{\rm
M}(\mref,y)=1/T(1+\frac12\sqrt{y},1-\frac12\sqrt{y})$. In order to
investigate the $y$-dependence numerically, we start from the
results for $a\mp$ and $a\Fp$ of UKQCD~\cite{impr:ukqcd_qhad} for
three values of the quark masses at each of the two bare couplings
$\beta=6.0,6.2$. Three combinations with $m_j \ne m_i$ 
are then available for each of the two corresponding
lattice spacings, $0.09\,\fm$ and $0.07\,\fm$. The only numerical
difficulty in examining the ratios $Q_X$ is that the denominators
$H_X(m,0)$ are in general not available directly. They may,
however, be replaced by an interpolation
\bes
 H_{\rm M}(m,0) &=& h^{\rm M}_1 m + h^{\rm M}_2 m^2 + h^{\rm M}_3 m^3\,, \\
 H_{\rm F}(m,0) &=& h^{\rm F}_0 + h^{\rm F}_1 m + h^{\rm F}_2 m^2\,,
\ees 
with $h_k^X$ determined from the observables computed for the three
available degenerate mesons ($m_i=m_j$). One has to check that these
interpolations are stable. This  was tested by taking different
subsets of three mass points of the data in~\tab{t_resABCD},
extracting $h_k^X$ and comparing the resulting functions $H_X(m,0)$. 
Deviations between these different interpolations were found to be
negligible compared to the statistical precision of the ratios $Q$
themselves.

Before discussing the results, let us mention one technical point in
the numerical analysis. For convenience, the above procedure was
carried out using the subtracted, improved bare quark masses,
$\mqtilde = \mq (1+\bm a \mq )\,,\;\mq = m_0 - \mc$, instead of the
current quark masses themselves.\footnote{Flavour indices are
suppressed, here, and we use $\bm=-1/2 - 0.0962 g_0^2$.}  For the
question addressed in this section, this makes no difference because
$ m = \zm \mqtilde $ holds up to the usual $\Oasq$ errors, with some
mass-independent renormalization factor $\zm$.

In \fig{f_Q} we show our numerical results.\footnote{The numerical
values for $(\mqtilde)_{\rm ref}$ were taken from the analysis
described in \sect{s_cont}.} They show that $Q_{\rm M}$ has a
dependence on $y$ which is below the level of a percent, and
$Q_{\rm F}$ is seen to be independent of $y$ to within our
statistical precision of around 0.3\%. It should be kept in mind
that~\fig{f_Q} tests the $y$-dependence for $m/\mref = $~1.4--2.4, while
later we shall be interested in kaon physics, where $m/\mref \approx
1$. It remains a (plausible!) hypothesis that $Q_X$ are close to
unity also for this lower value of $m$. Nevertheless, our
investigation quantifies the smallness of $y$-effects for the first
time and suggests that the correction terms in \eq{e_equiv} are
negligible. Based on this analysis we shall not distinguish between
$2\mref$ and $\mstrange +
\mlight$ from now on.

%% file: sect4.tex
\section{Finite size effects \label{s_fse}}

\input tab_resABCD.tex

Our numerical results, which are listed in \tab{t_resABCD}, have
been obtained for approximately constant physical volume, $T\times
L^3$, where $T=2L$ and $L\approx3\rnod=1.5\fm$. We used exactly the
numerical methods described in~\cite{mbar:pap2}, and in
particular systematic errors due to excited state contributions were
checked to be small compared to the statistical errors. Further
details about the simulation are described in the appendix. Since
the precision of the results is quite good one has to investigate
whether they may be affected by the finite size of the system at the
level of their statistical accuracy. The most relevant observable
for the determination of the quark masses is the ratio
$\Rwi=(m_i+m_j)/\mp^2$. Since $m_i$ is defined through the PCAC
relation, it is independent of the volume \cite{impr:pap1} (apart
from small lattice artefacts of order $a^2$). We are therefore
predominantly interested in the volume dependence of the {\ps}
masses but will also consider the decay constant $\Fp$.

In the {\ps} sector the leading finite size effects can be reliably
calculated in chiral perturbation theory. The reason is that for
vanishing quark masses the {\ps} mesons are Goldstone bosons. Their
interactions become weak for small energies and these are
responsible for the leading finite size effects for large but finite
volumes.

Gasser and Leutwyler have reported results for the two-point
function of the axial current at space-like separations
\cite{GaLe:87}. This result, given for finite temperature and in
finite volume is easily adapted to the situation which is of
interest here, namely the pseudoscalar mass on an $L^3$-torus
(temperature zero) $\mp(L)$, and the corresponding decay constant
$\Fp(L)$. The resulting formulae are
\bes
  {\mp(L) \over \mp(\infty)}-1 &=& {1 \over \nf} \,{\mp^2 \over \Fp^2} 
  \,g(z) + \rmO(\rme^{-\sqrt{2} z})\,, \;z=\mp\,L 
  \label{e_fse_chir1}
  \\
  {\Fp(L) \over \Fp(\infty)}-1 &=& -\,\nf  \,{\mp^2 \over \Fp^2} 
  \,g(z)+ \rmO(\rme^{-\sqrt{2} z})\,,
  \label{e_fse_chir2}
  \\ 
  g(z) &=& {3 \over 8\pi^2 \, z^2} \,\int_0^{\infty} {\rmd x \over x^{2}} 
                 \rme^{-z^2x - 1/(4x)} \,=\,
           {3 \over 2\pi^2 \, z} \, K_1(z) \,,
  \label{e_fse_chir3}
\ees
where $K_1(z)$ denotes a modified Bessel function. Here a comment is
in order. This is a result of the first non-trivial order in chiral
perturbation theory in full QCD. A priori one cannot expect it to be
accurate if the {\ps} meson masses are too large. On the other hand,
one knows that the finite size effects discussed here are of order
$\rme^{-z}$, as long as the {\ps} is the lightest particle in the
theory~\cite{FSE:martin1,FSE:martin2}. It is plausible that the
prefactor of the exponential is not dramatically different for
heavier {\ps} mesons. We may therefore take the above equations as a
reasonable estimate of finite size effects. Being interested in the
order of magnitude of these effects, we also do not worry about the
difference of the quenched approximation and full QCD.

In our numerical calculations (see Tables~\ref{t_resABCD}
and~\ref{tab_param}), we are in the range $z=\mp L \geq 4.3$ and
consequently eqs.~(\ref{e_fse_chir1})--(\ref{e_fse_chir2}) predict
corrections which are below the level of 0.5\% for $\Fp$ and 0.1\%
for $\mp$. In order to check this estimate, we have done some
calculations on lattices which are larger than our standard $L
\approx 1.5 \,\fm$. Entirely consistent with the above formulae, we
found no significant changes in $\Fp,\mp$, for instance
\vspace{1cm}
\bes
 \beta=6.2\,,\; a \mp = 0.208\,:&&
 {{\mp(24a)}\over{\mp(32a)}}-1 = 0.003(4)
  \;\;(\approx 0.0009)\,,
  \\
 &&{{\Fp(24a)}\over{\Fp(32a)}}-1 = -0.003(8)
  \;\;(\approx -0.003)\,,
\ees
where the numbers in parentheses are the estimates from the above equations
with $\nf=2$.

We conclude that even with the numerical precision
of~\tab{t_resABCD}, finite size effects are negligible for the quark
masses considered. Of course, as described by the above formulae,
finite size effects grow rapidly when the {\ps} mass becomes smaller
and larger volumes would be necessary for quark masses which are
significantly smaller than the ones we used.

%% file: tab_resABCD.tex
\begin{table}[tb]
\centering
\begin{tabular}{cr@{.}lr@{.}lr@{.}lr@{.}lr@{.}lr@{.}l}
\hline \\[-1.0ex]
$\beta$ & 
\multicolumn{2}{c}{$\kappa$} &
\multicolumn{2}{c}{$a m$} &
\multicolumn{2}{c}{$a m_{\rm PS}$} &
\multicolumn{2}{c}{$a m_{\rm V}$} &
\multicolumn{2}{c}{$aF_{\rm PS}$} &
\multicolumn{2}{c}{$F_{\rm PS} \over aG_{\rm PS}$}\\[1.0ex] 
\hline \\[-1.0ex]
6.0  & 0&1335  & 0&0466(1)  & 0&3884(10) & 0&5289(26) 
     & 0&0969(4) &  0&5997(31) \\
     & 0&1338  & 0&03856(6) & 0&3529(11) & 0&5077(31)
     & 0&0943(4) & 0&6022(34)  \\
     & 0&1340  & 0&03311(7) & 0&3275(11) & 0&4938(36)
     & 0&0924(4) & 0&6015(39)  \\
     & 0&1342  & 0&02759(8) & 0&3001(12) & 0&4804(44)
     & 0&0905(4) & 0&5978(45)  \\[1.0ex]
\hline \\[-1.0ex]
6.1  & 0&1342  & 0&04161(3) & 0&3306(5)  & 0&4542(17)
     & 0&0870(3) & 0&7454(22)  \\
     & 0&1345  & 0&03305(3) & 0&2947(5)  & 0&4328(23)
     & 0&0841(3) & 0&7476(26)  \\
     & 0&1347  & 0&02734(3) & 0&2687(6)  & 0&4188(28)
     & 0&0820(3) & 0&7458(30)  \\
     & 0&1349  & 0&02160(4) & 0&2399(7)  & 0&4056(39)
     & 0&0799(3) & 0&7405(37)  \\[1.0ex]
\hline \\[-1.0ex]
6.2  & 0&1347  & 0&03241(3) & 0&2683(5)  & 0&3748(21)
     & 0&0743(3) & 0&8887(33)  \\
     & 0&1349  & 0&02661(3) & 0&2430(6)  & 0&3588(26)
     & 0&0721(3) & 0&8908(38)  \\
     & 0&13515 & 0&01934(3) & 0&2080(6)  & 0&3388(39)
     & 0&0693(3) & 0&8857(48)  \\
     & 0&1352  & 0&01788(3) & 0&2004(6)  & 0&3348(43)
     & 0&0687(3) & 0&8830(51)  \\[1.0ex]
\hline \\[-1.0ex]
6.45 & 0&13485 & 0&02477(2) & 0&1975(8)  & 0&2749(25)
     & 0&0533(5) & 1&261(10)   \\
     & 0&1351  & 0&01734(2) & 0&1650(9)  & 0&2558(37)
     & 0&0503(4) & 1&265(14)   \\
     & 0&1352  & 0&01439(2) & 0&1505(9)  & 0&2480(46)
     & 0&0490(4) & 1&261(16)   \\
     & 0&1353  & 0&01142(2) & 0&1347(10) & 0&2398(59)
     & 0&0477(4) & 1&250(19)   \\[1.0ex]
\hline
\end{tabular}
\caption[]{\footnotesize
Results for masses and unrenormalized (ratios of) matrix elements
in all simulation points
\label{t_resABCD}
}
\end{table}

%% file: sect5.tex
\section{Degenerate quarks, continuum limit \label{s_cont}}
\begin{figure}[tb]
\hspace{0cm}
\vspace{-0.2cm}

\centerline{
\psfig{file=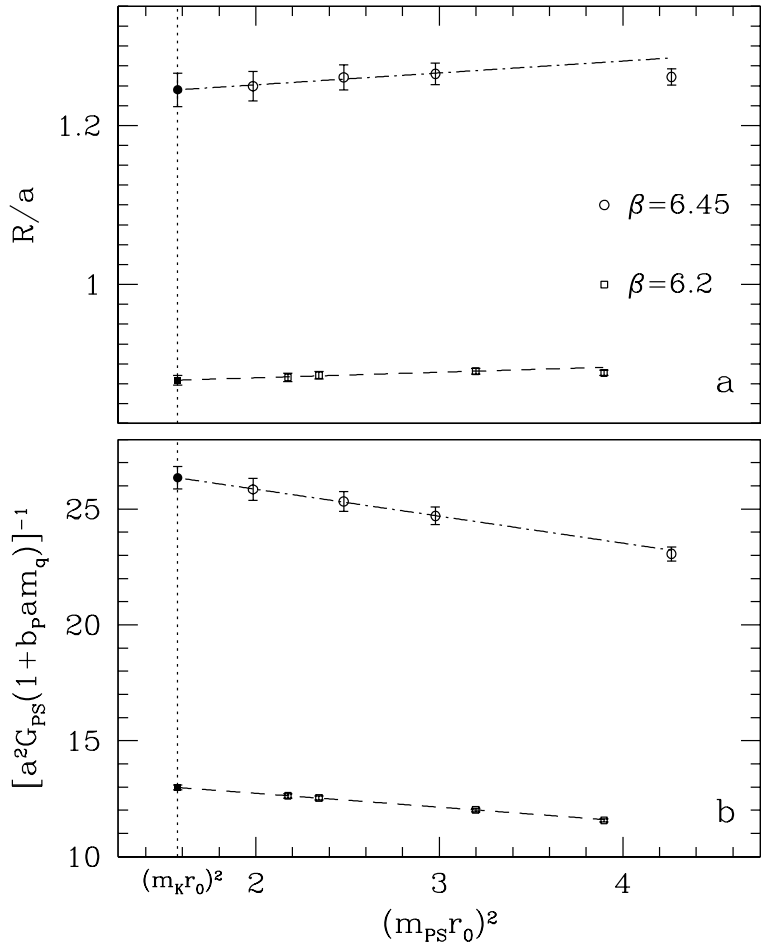,width=14cm}
}
\vspace{-1.0cm}
\caption{\footnotesize
Mass dependence and extrapolations for the two smallest values of the 
lattice spacing. The ratio $R$ is defined in
\eq{e_defR}.
\label{f_chir_ext_1}}
\end{figure}
%

We can now proceed to compute the desired quantity $r_0(M_{\strange}
+ \Mlight)$.
%
The first step is to evaluate $\Rwi$ for degenerate quarks as a
function of $r_0^2\mp^2$. Relegating all details of the exact
numerical procedure, such as the definition of $\mq$ and the
different improvement coefficients, to the appendix we directly show
the mass dependence of $\Rwi$ in \fig{f_chir_ext_1}a. It is apparent
that $\Rwi$ is almost constant as a function of the quark mass (or
$r_0^2\mp^2$). It is therefore easy to extrapolate to the desired
point $r_0^2\mp^2=r_0^2\mk^2=\rrmm$, slightly outside of the range
where we have numerical results. The extrapolation is performed
linearly in $r_0^2\mp^2$, using the three closest data
points. Simply taking for instance the closest data point or an
average of all of them would change the final result for the quark
mass by a negligible amount.

\input tab_extABCD.tex

\begin{figure}[tb]
\hspace{0cm}
\vspace{-0.2cm}

\centerline{
\psfig{file=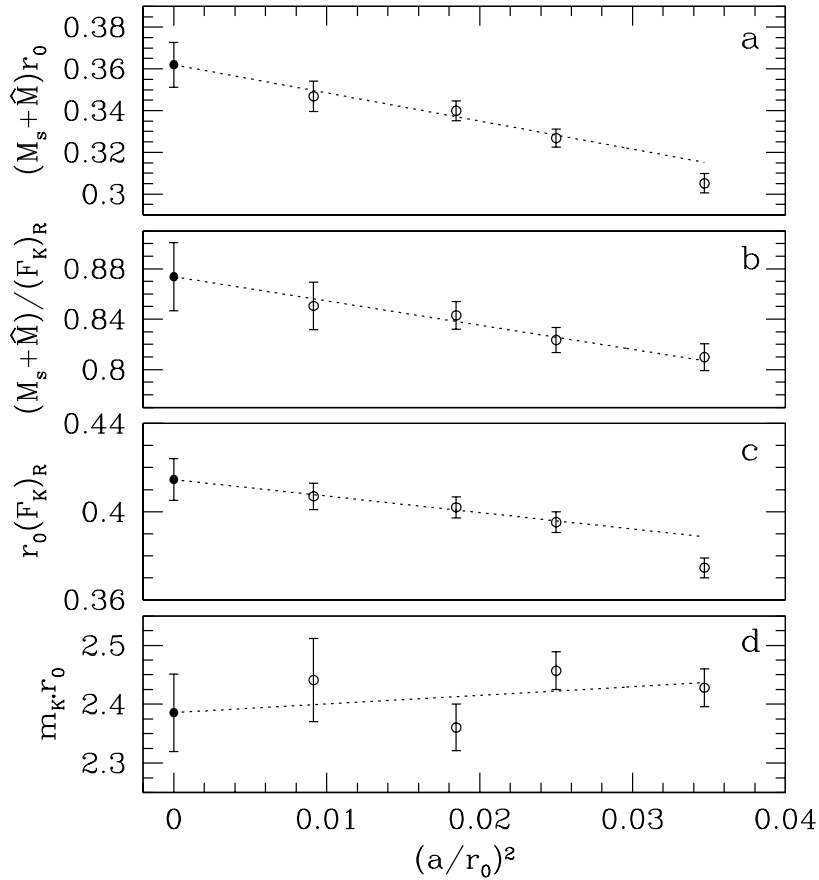,width=14cm}
}
\vspace{-1.2cm}
\caption{\footnotesize
Continuum limit extrapolations of several observables. Full symbols
show the extrapolated values. Dashed lines represent the
extrapolation function, which are continued outside the fit range
towards larger lattice spacings.
\label{f_cont_ext}}
\end{figure}

The second step is now to form the combination
\bes
  {r_0 (M_{\strange} + \Mlight) }
  = \zM {\Rwi|_{r_0^2\mp^2 =\rrmm} \over r_0} 
  \times \rrmm
\ees 
collecting all errors, including the ones on $\rnod$
\cite{pot:r0_SU3} and the $\beta$-dependent one of $\zM$
\cite{mbar:pap1}. We then extrapolate to the continuum limit
linearly in $a^2$. As seen in \fig{f_cont_ext}a, the dependence on
the lattice spacing is significant -- in contrast to the quark mass
on smaller volume \cite{impr:jochen}. As a safeguard against higher
order terms in the lattice spacing, we therefore include only the
three points with smallest lattice spacing in the extrapolation and
obtain our main result
\bes  
  r_0 (M_{\strange} + \Mlight) = 0.362(12)\,. 
\label{e_M_rnod}
\ees
This result contains the $\beta$-independent part of the uncertainty
in~$\zM$ of~1.3\% \cite{mbar:pap1}. At this point one may be
concerned about the validity of an $a^2$-extrapolation, since
$\ba-\bp$ is mostly known only in perturbation theory. However,
this combination is found to be tiny at the one-loop level
\cite{impr:pap5} and turns out to be rather small also
non-perturbatively~\cite{impr:roma2_1,impr:babp2}. We have checked
explicitly that such small values of $\ba-\bp$ do not affect our
continuum extrapolation.

In the whole procedure one may replace $\Rwi$, defined in
\eq{e_PCAC_l}, by the ratio of the current quark mass 
and the {\ps} mass squared. One then ends
up with slightly different lattice spacing effects and somewhat
different statistical errors. We have performed also that analysis,
using the current quark mass averaged over a range of timeslices
around the central one in our lattice. The results differ by up to
2\% from the ones presented above for finite lattice spacings but
are indistinguishable after the continuum extrapolation.

%

Instead of the quark mass in units of $\rnod$, we may also compute
the combination $(M_{\strange}+\Mlight)/(\Fk)_{\rm R}$, where
$(\Fk)_{\rm R}$ is the kaon decay constant. As its experimental
value we take $(\Fk)_{\rm R}=160(2)\,\MeV$. We note that
electromagnetic effects cannot be subtracted in analogy
to~\eq{e_m_K}, chiefly because $F_{\rm K^0}$ is not known
experimentally. We emphasize that this is truly an alternative
procedure which amounts to computing the combination
\bes
  {M_{\strange} + \Mlight \over (\Fk)_{\rm R}}
  = {M \over \mbar} 
  \left.{ 1 \over \zp \rnod^2 \Gp[1+ \bp a\mq]  }\right|_{r_0^2\mp^2 =\rrmm}
  \times \rrmm \label{e_M_F_def}\,,
\ees
where the ratio ${M / \mbar}$ and the renormalization constant $\zp$
have to be taken at a common scale for which we 
resort to the value $\mu=1/(1.436\rnod)$ chosen in
\cite{mbar:pap1}. Discretization effects are
clearly expected to be different in this case: they are known to be
non-negligible in the product $\rnod (\Fp)_{\rm R}$ \cite{lat97:hartmut}. A
non-perturbative estimate being unavailable at present, we have
to rely on the perturbative value $\bp=1 + 0.153\,g_0^2$
\cite{impr:pap5}. Extrapolations in $r_0^2\mp^2$ and the continuum
extrapolations are shown in \fig{f_chir_ext_1}b and
\fig{f_cont_ext}b. Concerning the former, the mass-dependence of 
${ 1/\{ \Gp[1+ \bp a\mq]\}}$ is stronger than the mass dependence of
$\Rwi$, but within our errors it is perfectly linear and the
extrapolation is easily done. In comparison to the quark mass in
units of $\rnod$, we find discretization errors which are roughly
only half as big.
 
In order to confirm that it is legitimate to perform an
extrapolation linear in $a^2$, even though the improvement
coefficient $\bp$ is known only perturbatively, we have repeated the
whole analysis after setting $\bp$ to its tree-level value
$\bp=1$. Results after extrapolation changed by much less than a
percent. Our final continuum result is
\bes
  {M_{\strange} + \Mlight \over (\Fk)_{\rm R}}
  = 0.874(29) \,.
 \label{e_M_F}
\ees
For illustration we may translate to physical units, setting
 $\rnod=0.5\,\fm$ \cite{pot:r0_SU3} and
$(\Fk)_{\rm R}=160(2)\MeV$. We thus obtain
\bes
  M_{\strange} + \Mlight &=& 143(5)\,\MeV \quad \mbox{from
    \eq{e_M_rnod}}\,,
 \label{e_M_Mev_rnod} \\
  M_{\strange} + \Mlight &=& 140(5)\,\MeV \quad \mbox{from \eq{e_M_F}}\,.
 \label{e_M_Mev_F}
\ees
As will be discussed in more detail below, this assignment of
physical units is ambiguous in the quenched approximation. One
should be well aware that the solid results are given in
\eq{e_M_rnod} and \eq{e_M_F}.

For future reference we list various dimensionless quantities both
for $\rnod^2\mp^2=\rrmm$ and for $\rnod^2\mp^2=3$
in~\tab{t_extABCD}. Our results before and after the continuum
extrapolation can be found in that table.

%% file: tab_extABCD.tex
\begin{table}[tb]
\centering
\begin{tabular}{llr@{.}lr@{.}lr@{.}lr@{.}lr@{.}l}
\hline \\[-1.0ex]
$r_0^2m_{\rm PS}^2$ & 
$\beta$ &
\multicolumn{2}{c}{$Z_{\rm M}\,\frac{R}{r_0}$} &
\multicolumn{2}{c}{$\frac{M_{\rm s}+\hat{M}}{(F_{\rm PS})_{\rm R}}$} &
\multicolumn{2}{c}{$r_0(F_{\rm PS})_{\rm R}$} &
\multicolumn{2}{c}{$m_{\rm V}r_0$} &
\multicolumn{2}{c}{$\frac{(F_{\rm PS})_{\rm R}}{m_{\rm V}}$}\\[1.0ex] 
\hline \\[-1.0ex]
1.5736 & 6.0  & 0&1939(30) & 0&810(11) & 0&3746(45) 
       & 2&428(32) & 0&1543(26) \\
       & 6.1  & 0&2077(28) & 0&824(10) & 0&3952(47) 
       & 2&457(32) & 0&1609(26) \\
       & 6.2  & 0&2160(30) & 0&843(11) & 0&4020(48) 
       & 2&361(40) & 0&1701(32) \\
       & 6.45 & 0&2205(46) & 0&851(19) & 0&4070(60) 
       & 2&441(71) & 0&1667(52) \\
       & {\it CL}
       & {\it 0}&{\it 2300(69)} 
       & {\it 0}&{\it 874(27)} 
       & {\it 0}&{\it 4146(94)} 
       & {\it 2}&{\it 386(66)} 
       & {\it 0}&{\it 1761(75)} \\[1.0ex]
\hline \\[-1.0ex]
3.0    & 6.0  & 0&1966(26) & 1&450(17) & 0&4063(47) 
       & 2&637(22) & 0&1541(20) \\
       & 6.1  & 0&2104(26) & 1&471(17) & 0&4288(49) 
       & 2&666(20) & 0&1609(20) \\
       & 6.2  & 0&2181(28) & 1&501(18) & 0&4363(51) 
       & 2&606(25) & 0&1675(23) \\
       & 6.45 & 0&2236(37) & 1&518(28) & 0&4421(62) 
       & 2&677(42) & 0&1652(32) \\
       & {\it CL}
       & {\it 0}&{\it 2324(57)} 
       & {\it 1}&{\it 553(42)} 
       & {\it 0}&{\it 4507(99)} 
       & {\it 2}&{\it 645(41)} 
       & {\it 0}&{\it 1709(49)} \\[1.0ex]
\hline
\end{tabular}
\caption[]{\footnotesize
Extra-/interpolations in $r_0^2m_{\rm PS}^2$ and resulting continuum
limits (CL) including all errors
\label{t_extABCD}
}
\end{table}

%% file: sect6.tex
\section{Ambiguities in the quenched approximation}

In the previous section, we have mainly used $\rnod$ as our
reference scale. However, in general one expects that by choosing
different experimental inputs one will also get somewhat different
results when working in the quenched approximation. Here we would
like to quantify this ambiguity for our determination of
$M_{\strange}+\Mlight$.

A first estimate is obtained by computing the combination
$\rnod(\Fk)_{\rm R}$, using our results for the pseudoscalar decay
constant including the non-perturbative renormalization factor
$\za$, and comparing it to its experimental value. Of course, on the
basis of the results for $M_{\strange}+\Mlight$ presented in
eqs.~(\ref{e_M_Mev_rnod}) and~(\ref{e_M_Mev_F}) one may not expect
any significant deviation. However, the following detailed analysis
may still be instructive, in particular since it yields
$\rnod(\Fk)_{\rm R}$ in the quenched approximation.

\begin{figure}[tb]
\hspace{0cm}
\vspace{-0.2cm}

\centerline{
\psfig{file=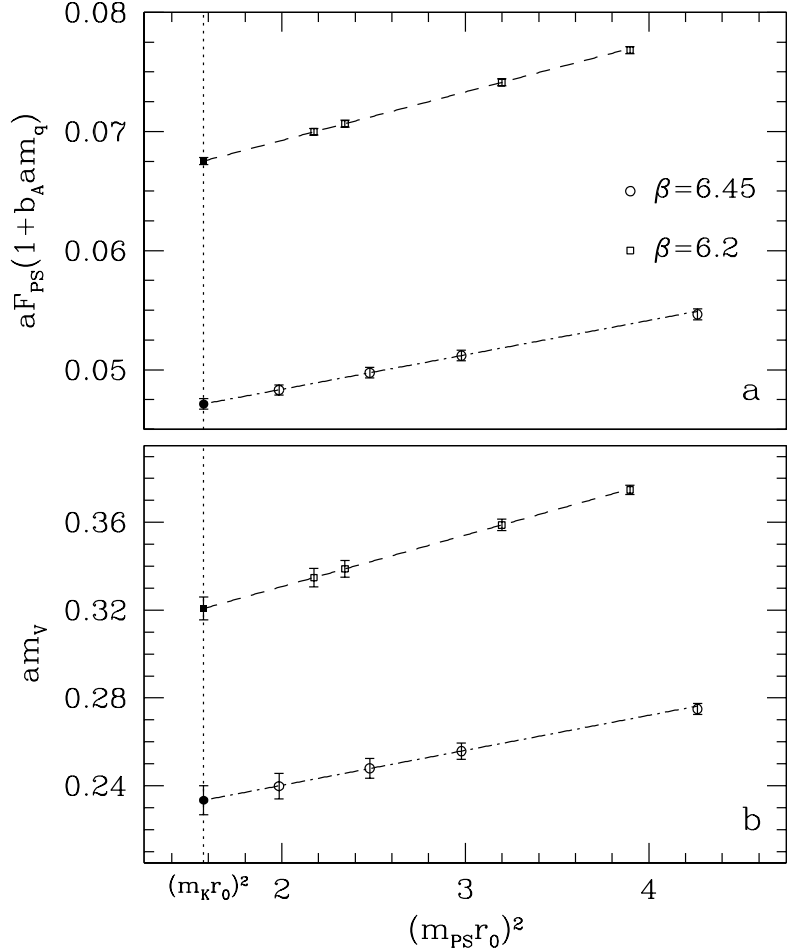,width=14cm}
}
\vspace{-1.2cm}
\caption{\footnotesize
Mass dependence of {\ps} decay constant and vector meson mass.
\label{f_chir_ext_2}}
\end{figure}

The procedure to extract $\rnod(\Fk)_{\rm R}$ is entirely analogous to what
was done in the previous section. The extrapolations in $r_0^2\mp^2$
(for mass-degenerate quarks) and the continuum extrapolations are
shown in \fig{f_chir_ext_2}a and \fig{f_cont_ext}c, respectively. In
the continuum limit we find
\bes
  \mbox{quenched:}\quad 
  \left\{\rnod \za \Fp [1+ \ba a\mq]\right\}_{r_0^2\mp^2 =\rrmm} =
  0.415(9)
\ees
compared to 
\bes
  \mbox{experiment:}\quad  0.5\,\fm \times (\Fk)_{\rm R} = 0.405(5)
\ees
with the experimental value of $(\Fk)_{\rm R}$. Clearly the result
obtained in the quenched approximation agrees well with experiment
when the approximate value $\rnod=0.5\,\fm$ is employed. However,
it would be premature to conclude that the ambiguity is small, since
the agreement may be special to the case of $(\Fk)_{\rm R}$.

%
Another alternative to set the scale is provided by the nucleon
mass, $m_{\rm N}$. Compared with other widely used quantities like
$m_\rho$, the nucleon has the advantage of being a stable
hadron. Although we have not computed $m_{\rm N}$ ourselves, we can
still use the precise published results obtained by the CP-PACS
Collaboration~\cite{qspect:CPPACS} and combine them with the values
of $r_0/a$ reported in~\cite{pot:r0_SU3}. We obtain
\bes
  \mbox{quenched:}\quad 
  \rnod m_{\rm N} \approx
  2.6
\ees
compared to 
\bes
  \mbox{experiment:}\quad  0.5\,\fm \times m_{\rm N} = 2.38 \,,
\ees
which represents a 10\% difference. Of course also the result for
$M_{\strange}+\Mlight$ in MeV would change by the same amount if
one used $m_{\rm N}$ instead of $\rnod$ or $(\Fk)_{\rm R}$ as experimental
input. This may serve as a rough estimate of the inherent ambiguity
in the quenched model.

In principle one could also use directly the results from recent
comparisons of the quenched light hadron spectrum with experiment
\cite{qspect:GF11,qspect:CPPACS} to estimate this ambiguity. In
ref.~\cite{qspect:CPPACS} statistically significant deviations of up 
to 11\% are
observed when the scale is set by $m_\rho$, which, as a resonance,
has a fairly large width. Since we consider it
safer to set the scale by the
nucleon mass, we have refrained from quoting this number as the typical
uncertainty for the case at hand. However, using the published
numbers of \cite{qspect:CPPACS}, it is not difficult to estimate the
quenched results for the other hadron masses when one chooses
$m_{\rm N}$ as input -- at least within a precision of
2-3\%. Interestingly it then turns out that the masses of the stable
hadrons agree with experiment to within about~4\%. On the other
hand, for unstable hadrons one observes differences which can be as
large as their widths. This may not be too surprising, since
resonance effects are not controlled in the lattice calculation.
In short, it is difficult to assess the relevance of the deviations
observed in the quenched hadron spectrum for our analysis, and thus
we stick to our above estimate of around 10\% for the ambiguity in question.

Finally, let us discuss the quark mass dependence of the {
flavour non-singlet} vector meson masses in more detail. First we
note that effects of the differences of quark masses can be shown to
be unimportant in the same way as in \sect{s_ndg} and we restrict
our attention to the masses of mass-degenerate mesons
($m_i=m_j$). Our aim is to map out the quark mass dependence of
$\mv$ in the continuum limit: we pick certain values of $q\equiv
m/\m_\strange=M/\Mstrange$, and for each of these values and for
each lattice spacing we then perform an inter-/extrapolation of
$\mv$ as a function of the quark mass to determine
$\mv(m=q\,m_\strange)$. Here, $m_\strange = 1.921\, \mref$
(see~\eq{e_ratios}) is used and $\mref$ is known from the previous
section. At fixed value of $q$ we then extrapolate $\rnod \mv$ to
the continuum limit including our data for all values of the lattice
spacing, after observing that the lattice spacing dependence is very
weak.

\begin{figure}[tb]
\hspace{0cm}
\vspace{-3.7cm}

\centerline{
\psfig{file=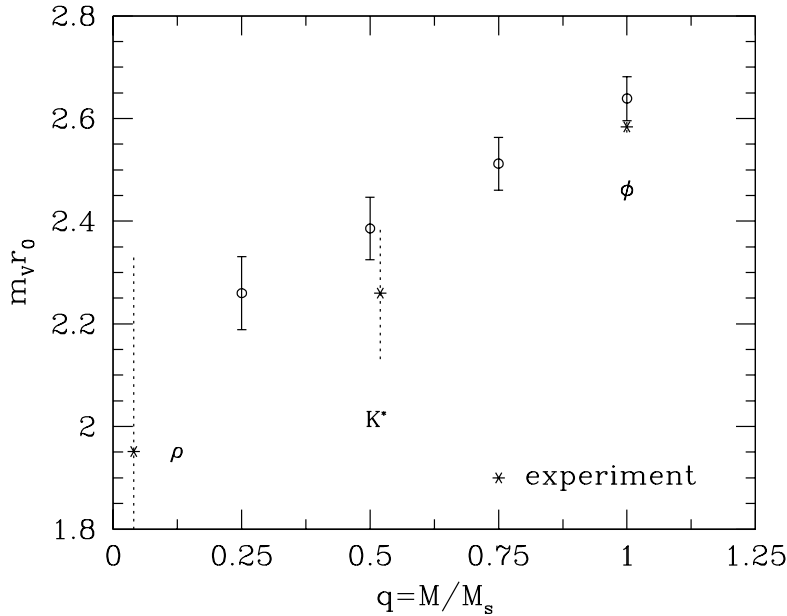,width=14cm}
}
\vspace{-2.2cm}
\caption{\footnotesize
Quark mass dependence of flavour non-singlet vector meson masses.
Experimental masses are shown as asterisks and their width is indicated 
by a dotted line. 
\label{f_cont_mVz}}
\end{figure}

In \fig{f_cont_mVz}, we plot the continuum results as a function of
$q$. They are compared to $\mkstar$, $m_\rho$ and $m_\phi$. Since
all of these states are resonances, we also indicate their width in
the figure. Additional reservations apply to the inclusion of the
$\phi$ meson in this comparison. It is a mixture of flavour octet
{\it and} singlet components, while the lattice calculation is for a
pure octet state with mass-degenerate quarks: the disconnected quark
diagrams occurring in the singlet channel are not accounted for, and
mixing with glueballs is neglected. Although all these effects may
be argued to be irrelevant in the quenched approximation, the
experimental value in \fig{f_cont_mVz} does include them and it is
somewhat surprising that our values for $\mv$ at $q=1$ come so close
to $m_\phi$.

If we were to ignore these problems and determined $\Mstrange$ from
the requirement that $\rnod \mv$ at that quark mass agrees with
$\rnod m_\phi$, it is evident from \fig{f_cont_mVz} that we would
obtain a result for $\Mstrange$ which is compatible with the numbers
quoted in the previous section. However, the error would be
considerably larger. Further continuum results such as $\rnod
\mkstar$, shown also in \fig{f_chir_ext_2}b and
\fig{f_cont_ext}d, and $(\Fk)_{\rm R}/ \mkstar$ are listed in
\tab{t_extABCD} for future reference.

To summarize, we have directly seen ambiguities of about 10\% owing
to the use of the quenched approximation. This has to be kept in
mind when quark masses in $\MeV$ are quoted. Still we have
demonstrated that within the quenched approximation a total
precision of 3\% (see~\eq{e_M_rnod} and~\eq{e_M_F}) can be achieved.

%% file: sect7.tex
\section{Discussion}
Noting that ratios of light quark masses can at present best be calculated
in chiral perturbation theory, we discussed a strategy to
compute the overall scale of the light quark masses and
applied it to the test-case of quenched QCD.

The problem has essentially two parts. First, the renormalization
problem should be solved non-perturbatively. Here we have used the
recent solution of the
ALPHA-collaboration~\cite{mbar:pert,mbar:pap1} which connects the
bare current quark masses to the renormalization group invariant
ones. Second, one particular quark mass should be computed by
matching with {\ps} meson masses. It is not necessary to perform
this matching at physical values of the quark masses, because the
parameters in the effective chiral Lagrangian are
mass-independent. For both the quenched approximation and full QCD
it appears convenient to use two different (flavours of) quarks with
equal mass $\Mref$ such that $\mp$ is equal to the kaon mass $\mk$.
 
The ratio of $\Mref$ to the physical quark masses can then be
computed in chiral perturbation theory. In \sect{s_ndg} we have
shown that {\ps} masses, $\mp$, have {\em very little} dependence on
the difference of the two quark masses in the quenched
approximation. From this one has $ 2\,\Mref
\approx\Mstrange+\Mlight, $ which is also true in chiral
perturbation theory (in full QCD!).

By making the identification $2\Mref=\Mstrange+\Mlight$ we have
computed $(\Mstrange+\Mlight)\rnod$ and
$(\Mstrange+\Mlight)/(\Fk)_{\rm R}$, incorporating all systematic
errors (such as finite size effects), and have taken the continuum
limit of these quantities. These solid results are given in
\eq{e_M_rnod} and
\eq{e_M_F}.

A conversion of the results to physical units is necessarily
ambiguous, since quenched QCD does not describe the real
world. Depending on the choice of quantity somewhat different
results in~MeV are obtained. It is interesting to note that using
either $\rnod=0.5\,\fm$ or $(\Fk)_{\rm R}=160\,\MeV$ gives indistinguishable
results at our level of precision of 3\%. However, this does
not mean that $\Mstrange+\Mlight$ has to be very close to our
quenched result also in full QCD. Indeed, we have estimated that
replacing $(\Fk)_{\rm R}$ by $m_{\rm N}$ (or other stable light hadron masses)
would give quark masses in MeV, which are larger by about 10\%.

\input tab_mbarMSbar.tex

Usually the running quark masses in the $\msbar$ scheme,
$\mbar_\msbar(\mu)$, are quoted. Their relation to the
renormalization group invariant quark masses, which we computed here,
has been discussed in Sect.~2 of \cite{mbar:pap1}. For convenience,
we include a table with the perturbative conversion factors
$\mbar_\msbar(\mu)/M$ computed by numerical integration of the
renormalization group equations with the $n$-loop approximations of
the renormalization group functions ($\beta,\tau$ in the notation of
\cite{mbar:pap1}) for $n=2,3,4$. One should be aware that the result
$\Lambda_{\msbar}^{(0)} = 238(19)\,\MeV$ \cite{mbar:pap1} enters the
numbers of~\tab{t_mbarMSbar}. The uncertainty in $\Lambda$
translates into $1.5\%$ in $\mbar_\msbar(\mu)/M$ at $\mu=2\,\GeV$
and $2.5\%$ at $\mu=1\,\GeV$. If desired, \tab{t_mbarMSbar} may be
used to estimate $\mbar_\msbar(\mu)$. A typical result combining
\eq{e_M_Mev_F}, \eq{e_ratios} and the table is
\bes
  \mbar_{\strange}(2\,\GeV) =97(4)\,\MeV \quad
  \mbox{with 4-loop running in the $\MSbar$ scheme}\,.
\ees
This illustrates that a high level of precision can be reached by
state of the art lattice techniques.

Effects of dynamical fermions may first be examined in the theory
with unphysically large quark masses. We have also given results for
a reference mass which is roughly twice the strange quark mass. One
should quantify what the effects of dynamical fermions are at such a
point, where a good accuracy may be achieved, and then move on to the
more chiral regime.

Also for some quantities not related directly to quark masses,
precise results have been obtained. They refer to the continuum
limit of the quenched approximation and are summarized in
\tab{t_extABCD}. Most notably, decay constants were computed with a
precision not much worse than the experimental one. This was
achieved using the method to compute hadronic correlation functions
proposed in~\cite{mbar:pap2}. In this context we should also
comment on the continuum extrapolations in the $\Oa$ improved
theory. They are certainly necessary! For the quark mass in units
of $\rnod$, the difference between its value at a lattice spacing
$a=0.1\,\fm$ and at $a=0$ amounts to about 15\%. Other quantities
show smaller lattice spacing effects as was observed also in a
finite volume study \cite{impr:jochen}. Although clearly one would
have hoped for a weaker $a$-dependence, we note that their order of
magnitude is not much different from what is known already for pure
gauge theory observables \cite{pot:r0_SU3}.

Finally let us come back to the r\^ole of lattice QCD in determining
parameters of the chiral Lagrangian. Figs. \ref{f_chir_ext_1} and 
\ref{f_chir_ext_2}a show to what
precision the quark mass dependence of observables in the
pseudoscalar sector can be computed. Such information, once
available in full QCD, will allow to reduce the errors in the chiral
Lagrangian. For instance, we have checked that just the mass
independence of $R$ puts strong restrictions on these parameters. As
usual, estimates of the uncertainties is a delicate issue and a
detailed investigation of the potential of this approach will be
left for future work.

%% file: tab_mbarMSbar.tex
\begin{table}[tb]
\centering
\begin{tabular}{cr@{.}lr@{.}lr@{.}l}
\hline \\[-1.0ex]
$\mu\,\,[\,{\rm GeV}\,]$ & \multicolumn{6}{c}{$\mbar_{\msbar}(\mu)/M$}
\\[1.0ex] 
& \multicolumn{2}{c}{2-loop} &
\multicolumn{2}{c}{3-loop} &
\multicolumn{2}{c}{4-loop} \\[1.0ex] 
\hline \\[-1.0ex]
   1.0 & 0&80279 & 0&83585 & 0&84449 \\
   2.0 & 0&70388 & 0&71830 & 0&72076 \\
   4.0 & 0&64079 & 0&64880 & 0&64981 \\
   8.0 & 0&59549 & 0&60055 & 0&60105 \\
  90.0 & 0&49937 & 0&50105 & 0&50112 \\[1.0ex]
\hline
\end{tabular}
\caption[]{\footnotesize
Factors to convert the renormalization group invariant mass into
the $\msbar$ scheme at scale $\mu$ 
\label{t_mbarMSbar}
}
\end{table}

%% file: ack.tex
\vspace{1cm}
The project of computing light quark masses was started quite a
while ago by the ALPHA collaboration. Here we presented the final
results.
The basis of our approach was developed together with M. L\"uscher,
S. Sint and P. Weisz. We would like to thank them for a most
enjoyable collaboration.
In particular, R.S. and H.W. thank
M. L\"uscher for numerous enlightening discussions. Furthermore, we
appreciate useful correspondence with G. Ecker and D. Wyler on
chiral perturbation theory and electromagnetic corrections.\\ The
numerical computations were performed on the APE100/Quadrics
computers of DESY Zeuthen and on the Cray T3D at EPCC Edinburgh. We
thank these institutions for their support. This work was in part
supported by EPSRC grant GR/K41663 and PPARC grants GR/K55745 and
GR/L29927.

%% file: appendix1.tex
\section{Numerical details}
\subsection{Improvement coefficients, renormalization factors}

Our calculations have been performed using the $\Oa$ improved Wilson 
action defined in ref.~\cite{impr:pap3}, which can be consulted for
any unexplained notation. In particular, the improvement coefficients 
$\csw$ and $\ca$ were taken from eqs.~(5.15) and~(6.4) of that
reference. 

The renormalized axial current and pseudoscalar density for quark
flavours~$i$ and~$j$ are defined as
\bes
(\ar)_\mu(x) &=& \za(1+\ba a\mq)(\aimpr)_\mu(x),  \\
 \pr(x)      &=& \zp(1+\bp a\mq)P(x),
\ees
where $P(x)=\psibar_i(x)\gamma_5\psi_j(x)$, and the improved,
unrenormalized axial current is given by
\bes
  (\aimpr)_\mu(x) = \psibar_i(x)\gamma_\mu\gamma_5\psi_j(x)
  +a\ca\frac{1}{2}\left(\partial^*_\mu+\partial_\mu\right) P(x).
\label{e_axialimpdef}
\ees
The renormalization factor $\za$ was calculated in \cite{impr:pap4}, 
and here we have used its parameterization in eq.~(6.11) of that
paper. 

The renormalization factor $\zM$, which relates the current quark
mass in the $\Oa$ improved theory to the renormalization group
invariant quark mass, has recently been determined~\cite{mbar:pap1}.
In that paper also the scale- and scheme-dependent factor $\zp$ has
been computed in the Schr\"odinger functional (SF) scheme at a fixed
scale of $L=1.436\,r_0$. Here we use their representation in terms
of polynomial fit functions, viz.
\bes
& &\zM(g_0) = 1.752 +0.321\,(\beta-6) -0.220\,(\beta-6)^2\,,\\
& &\zp(g_0,L/a)_{L=1.436\,r_0} = 0.5233-0.0362\,(\beta-6)
                                +0.0430\,(\beta-6)^2\,, \\
& &\beta=6/g_0^2,\qquad 6.0\leq\beta\leq6.5. \nonumber
\label{e_zMzp_param}
\ees
As explained in Subsect.~6.2 of ref.~\cite{mbar:pap1}, the
uncertainty in $\zM$ is split into a $\beta$-dependent part of
1.1\%, which enters any continuum extrapolation, and a
$\beta$-independent error of 1.3\% which must be added (in
quadrature) to the extrapolated result. The typical accuracy of
$\zp$ as given by the above polynomial is 0.5\%.

The factor $\zM$ renormalizes the lattice counterpart of the ratio
$R(m_i,m_j)$ defined in eq.~(\ref{e_def_R}). In the $\Oa$ improved
theory it is given by
\bes
&  & R(m_i,m_j) = {\Fp \over \Gp}\left[1+(\ba-\bp)a\mq\right]\,, 
  \label{e_defR}
\ees
where
\bes
  \mq = \frac{1}{2}\left\{(\mq)_i+(\mq)_j\right\}=
  {1\over4a}
  \left({1\over\kappa_i}+{1\over\kappa_j}-{2\over\hopc}\right)\,,
\ees
and $\kappa_i,\,\kappa_j$ are the hopping parameters of flavours~$i$ 
and~$j$.

The combination $\ba-\bp$ has been found to be small in perturbation
theory~\cite{impr:pap5} and also non-perturbatively
\cite{impr:roma2_1,impr:babp2}. A method how to determine the individual
coefficients~$\ba$ and~$\bp$ non-perturbatively has been proposed in
ref.~\cite{impr:rajan_etal2}, but no results for our choice of
action have been reported so far.

The values for the critical hopping parameter $\hopc$ have been
taken from Table~1 of ref.~\cite{impr:pap3}. When necessary they
have been interpolated linearly to the desired $\beta$-value.

\subsection{Hadronic correlation functions, reference scale}

Our calculation of hadronic correlation functions using
Schr\"odinger functional boundary conditions follows exactly the
procedures outlined in a previous paper~\cite{mbar:pap2}. In
particular, this reference contains the definitions of the
correlation functions $\fp,\,\faimpr,\,\kvimpr$ and $f_1$, which we
have computed to extract hadron masses and matrix elements for
pseudoscalar and vector mesons.

Estimates for meson masses were obtained by averaging effective
masses, 
\bes
  am_{{\rm eff},X}(x_0+\frac{a}{2}) = \ln\left(f(x_0)/f(x_0+a)\right)\,,
  \quad  { X}={\rm PS,\,V},
\ees
over a suitably chosen interval $\tmin\leq x_0\leq\tmax$.
Similarly, the ratio $\Fp/\Gp$ was obtained by averaging the
combination 
\bes
{\Fp\over\Gp} = {1\over\mp}{\faimpr(x_0)\over\fp(x_0)},
\ees
and the individual matrix elements $\Fp$ and $\Gp$ are extracted from 
\bes
  \Fp &=& 2(\mp\, L^3)^{-1/2} \rme^{(x_0-T/2)\mp} \,
   {\faimpr(x_0) \over \sqrt{f_{1}}} \,,  \\
  \Gp &=& 2(\mp/L^3)^{1/2} \rme^{(x_0-T/2)\mp} \,
   {\fp(x_0) \over \sqrt{f_{1}}} \,,
\ees
where again an average over the interval $[\tmin,\tmax]$ is
taken. The averaging intervals used here were the same as those in
Table~1 of ref.~\cite{mbar:pap2}. 

We have also calculated an unrenormalized current quark mass
defined by
\bes
   m(x_0) = {\frac{1}{2}(\partial_0^*+\partial_0)\fa(x_0)
   +{\ca}a\partial_0^*\partial_0\fp(x_0) \over 2\fp(x_0)}\;.
\ees
The values for the current quark mass listed in \tab{t_resABCD} have 
been obtained by averaging $m(x_0)$ over the interval
$T/4\leq{x_0}\leq3T/4$.

For the hadronic radius $r_0$ which was used to set the scale, we
have taken its parameterization quoted in \cite{pot:r0_SU3}, viz.
\bes
  \ln(a/r_0) = -1.6805  -1.7139\,(\beta-6)
  +0.8155\,(\beta-6)^2 -0.6667\,(\beta-6)^3,
\label{e_poly_fit}
\ees
accounting for the uncertainty as quoted in  \cite{pot:r0_SU3}.
In addition to the results obtained at $\beta=6.0$ and~6.2, which have 
been published in~\cite{mbar:pap2}, we have also computed
correlation functions at $\beta=6.1$ and $6.45$. A brief summary of
our simulation parameters including $N_{\rm
meas}$, the number of gauge field configurations for which
the correlation functions were computed,
 is presented in Table~\ref{tab_param}.

\begin{table}[htb]
\centering
\begin{tabular}{l c c r@{.}l r}
\hline \\[-1.0ex]
$\beta$ & $T/a$ & $L/a$ & \multicolumn{2}{c}{$L/r_0$}
 & $N_{\rm meas}$ \\[1.0ex]
\hline \\[-1.0ex]
 6.0  & 32 & 16 &  2&981(12)  & 1000  \\
 6.1  & 40 & 24 &  3&795(17)  &  800  \\
 6.2  & 48 & 24 &  3&261(16)  &  800  \\
 6.45 & 64 & 32 &  3&060(17)  &  220  \\[1.0ex]
\hline
\end{tabular}
\caption[tab_param]{\footnotesize Simulation parameters  \label{tab_param}}
\end{table}

%% file: pap3.bbl
\begin{thebibliography}{10}

\bibitem{reviews:quarkmasses}
J. Gasser and H. Leutwyler,
\newblock Phys. Rept. 87 (1982) 77.

\bibitem{leutwyler:94}
H. Leutwyler,
\newblock (1994), hep-ph/9406283.

\bibitem{leutwyler:1996}
H. Leutwyler,
\newblock Phys. Lett. B378 (1996) 313, hep-ph/9602366.

\bibitem{quark:APE1}
C.R. Allton et~al.,
\newblock Nucl. Phys. B431 (1994) 667, hep-ph/9406263.

\bibitem{quark:APE2}
C.R. Allton, V. {Gim\'enez}, L. Giusti and F. Rapuano,
\newblock Nucl. Phys. B489 (1997) 427, hep-lat/9611021.

\bibitem{quark:APE3}
V. {Gim\'enez}, L. Giusti, F. Rapuano and M. Talevi,
\newblock Nucl. Phys. B540 (1999) 472, hep-lat/9801028;
L. Giusti, V. {Gim\'enez}, F. Rapuano, M. Talevi and A. Vladikas,
\newblock (1998), hep-lat/9809037.

\bibitem{hadr:ukqcd}
UKQCD Collaboration, C.R. Allton et~al.,
\newblock Phys. Rev. D49 (1994) 474, hep-lat/9309002.

\bibitem{quark:gupta}
R. Gupta and T. Bhattacharya,
\newblock Phys. Rev. D55 (1997) 7203, hep-lat/9605039.

\bibitem{quark:gough}
B.J. Gough, G.M. Hockney, A.X. El-Khadra, A.S. Kronfeld,
P.B. Mackenzie, B.P. Mertens, T. Onogi and J.N. Simone,
\newblock Phys. Rev. Lett. 79 (1997) 1622, hep-ph/9610223.

\bibitem{impr:qcdsf}
M. {G\"ockeler}, R. Horsley, H. Perlt, P. Rakow, G. Schierholz,
A. Schiller and P. Stephenson,
\newblock Phys. Rev. D57 (1997) 5562, hep-lat/9707021.


\bibitem{impr:roma2}
A. Cucchieri, M. Masetti, T. Mendes and R. Petronzio,
\newblock Phys. Lett. B422 (1998) 212, hep-lat/9711040;
A. Cucchieri, T. Mendes and R. Petronzio,
\newblock J. High Energy Phys. 05 (1998) 006, hep-lat/9804007.

\bibitem{quark:marti}
D. Becirevic, Ph. Boucaud, J.P. Leroy, V. Lubicz,
G. Martinelli and F. Mescia,
\newblock Phys. Lett. B444 (1998) 401, hep-lat/9807046.

\bibitem{quark:SESAM}
SESAM Collaboration, N. Eicker et~al.,
\newblock Phys. Lett. B407 (1997) 290, hep-lat/9704019;
SESAM Collaboration, N. Eicker et~al.,
\newblock Phys. Rev. D59 (1999) 014509, hep-lat/9806027.

\bibitem{quark:dwf}
T. Blum, A. Soni and M. Wingate, 
\newblock (1999), hep-lat/9902016.

\bibitem{quark:jlqcd_ks}
JLQCD Collaboration, S. Aoki et al., 
\newblock Phys. Rev. Lett. 82 (1999) 4392, hep-lat/9901019.

\bibitem{qspect:CPPACS}
CP-PACS Collaboration, S. Aoki et al.,
\newblock (1999), hep-lat/9902018.

\bibitem{quark:bijnens}
J. Bijnens, J. Prades and E. de~Rafael,
\newblock Phys. Lett. B348 (1995) 226, hep-ph/9411285.

\bibitem{quark:jamin1}
M. Jamin and M. {M\"unz},
\newblock Z. Phys. C66 (1995) 633, hep-ph/9409335.

\bibitem{quark:jamin2}
M. Jamin,
\newblock Nucl. Phys. B (Proc. Suppl.) 64 (1998) 250, hep-ph/9709484.

\bibitem{quark:narison}
S. Narison,
\newblock Phys. Lett. B358 (1995) 113, hep-ph/9504333.

\bibitem{quark:chetyrkin}
K.G. Chetyrkin, D. Pirjol and K. Schilcher,
\newblock Phys. Lett. B404 (1997) 337, hep-ph/9612394.

\bibitem{quark:colangelo}
P. Colangelo, F.D. Fazio, G. Nardulli and N. Paver,
\newblock Phys. Lett. B408 (1997) 340, hep-ph/9704249.

\bibitem{quark:prades}
J. Prades,
\newblock Nucl. Phys. B (Proc. Suppl.) 64 (1998) 253, hep-ph/9708395.

\bibitem{quark:yndurain}
F.J. {Yndur\'ain},
\newblock Nucl. Phys. B517 (1998) 324, hep-ph/9708300.

\bibitem{quark:dosch}
H.G. Dosch and S. Narison,
\newblock Phys. Lett. B417 (1998) 173, hep-ph/9709215.

\bibitem{quark:lellouch}
L. Lellouch, E. de~Rafael and J. Taron,
\newblock Phys. Lett. B414 (1997) 195, hep-ph/9707523.

\bibitem{quark:maltman}
K. Maltman, 
           (1999),   hep-ph/9904370.

\bibitem{mbar:pap1}
ALPHA Collaboration, S. Capitani, M. {L\"uscher}, R. Sommer
and H. Wittig,
\newblock Nucl. Phys. B544 (1999) 669, hep-lat/9810063.

\bibitem{mbar:pert}
ALPHA Collaboration, S. Sint and P. Weisz,
\newblock Nucl. Phys. B545 (1999) 529, hep-lat/9808013.

\bibitem{Dashen}
R. Dashen,
\newblock Phys. Rev. 183 (1969) 1245.

\bibitem{em_corr}
K.~Maltman and D.~Kotchan,
\newblock Mod. Phys. Lett. A5 (1990) 2457;
J.F. Donoghue, B.R. Holstein and D. Wyler,
\newblock Phys. Rev. D47 (1993) 2089;
J.~Bijnens,
\newblock Phys. Lett. B306 (1993) 343, hep-ph/9302217;
R.~Baur and R.~Urech, 
\newblock Phys. Rev. D53 (1996) 6552, hep-ph/9508393;
J.F.~Donoghue,
\newblock private communication through D. Wyler.

\bibitem{pot:r0}
R. Sommer,
\newblock Nucl. Phys. B411 (1994) 839, hep-lat/9310022.

\bibitem{chir:GaLe1}
J. Gasser and H. Leutwyler,
\newblock Ann. Phys. 158 (1984) 142.

\bibitem{chir:GaLe2}
J. Gasser and H. Leutwyler,
\newblock Nucl. Phys. B250 (1985) 465.

\bibitem{chir:param}
J. Bijnens, G. Ecker and J. Gasser,
\newblock (1994), hep-ph/9411232.

\bibitem{impr:pap1}
M. {L\"uscher}, S. Sint, R. Sommer and P. Weisz,
\newblock Nucl. Phys. B478 (1996) 365, hep-lat/9605038.

\bibitem{impr:pap3}
M. {L\"uscher}, S. Sint, R. Sommer, P. Weisz and U. Wolff,
\newblock Nucl. Phys. B491 (1997) 323, hep-lat/9609035.

\bibitem{chir:quenched1}
S.R. Sharpe,
\newblock Phys. Rev. D41 (1990) 3233;
J.N. Labrenz and S.R. Sharpe,
\newblock Phys. Rev. D54 (1996) 4595, hep-lat/9605034.

\bibitem{chir:quenched2}
C.W. Bernard and M.F.L. Golterman,
\newblock Phys. Rev. D46 (1992) 853, hep-lat/9204007;
\newblock Nucl. Phys. B (Proc. Suppl.) 30 (1993) 217.

\bibitem{chir:quenched3}
M. Booth, G. Chiladze and A.F. Falk,
\newblock Phys. Rev. D55 (1997) 3092, hep-ph/9610532.

\bibitem{except:fermilab}
W. Bardeen, A. Duncan, E. Eichten, G. Hockney and H. Thacker,
\newblock Phys. Rev. D57 (1998) 1633, hep-lat/9705008.

\bibitem{impr:ukqcd_qhad}
UKQCD Collaboration, K.C. Bowler et al., 
\newblock Quenched QCD with O($a$) improvement: 'I. The spectrum of
light hadrons', in preparation.

\bibitem{mbar:pap2}
ALPHA Collaboration, M. Guagnelli, J. Heitger, R. Sommer and H. Wittig,
\newblock (1999), hep-lat/9903040.

\bibitem{GaLe:87}
J. Gasser and H. Leutwyler,
\newblock Phys. Lett. 184B (1987) 83.

\bibitem{FSE:martin1}
M. {L\"uscher},
\newblock Lectures given at Summer School 'Fields, Strings and Critical
  Phenomena', Les Houches, France, Jun 28 - Aug 5, 1988.

\bibitem{FSE:martin2}
M. {L\"uscher},
\newblock Commun. Math. Phys. 104 (1986) 177.

\bibitem{pot:r0_SU3}
ALPHA Collaboration, M. Guagnelli, R. Sommer and H. Wittig,
\newblock Nucl. Phys. B535 (1998) 389, hep-lat/9806005.

\bibitem{impr:jochen}
ALPHA Collaboration, J. Heitger,
\newblock (1999), hep-lat/9903016.

\bibitem{impr:pap4}
M. {L\"uscher}, S. Sint, R. Sommer and H. Wittig,
\newblock Nucl. Phys. B491 (1997) 344, hep-lat/9611015.

\bibitem{impr:roma2_1}
G.M. de~Divitiis and R. Petronzio,
\newblock Phys. Lett. B419 (1998) 311, hep-lat/9710071.

\bibitem{impr:babp2}
T. Vergata and ALPHA Collaborations,
\newblock work in progress.

\bibitem{lat97:hartmut}
H. Wittig,
\newblock Nucl. Phys. B (Proc. Suppl.) 63 (1998) 47, hep-lat/9710013.

\bibitem{impr:pap5}
S. Sint and P. Weisz,
\newblock Nucl. Phys. B502 (1997) 251, hep-lat/9704001.

\bibitem{qspect:GF11}
F. Butler, H. Chen, J. Sexton, A. Vaccarino and D. Weingarten,
\newblock Nucl. Phys. B430 (1994) 179, hep-lat/9405003.

\bibitem{impr:rajan_etal2}
T. Bhattacharya, S. Chandrasekharan, R. Gupta, W. Lee and S. Sharpe,
\newblock (1999), hep-lat/9904011.

\end{thebibliography}
